\documentstyle[prd,aps,preprint,tighten,epsfig]{revtex}
\begin{document}
\draft
\title{Relating the neutrino mixing angles to a lepton mass hierarchy}
\author{{\bf Harald Fritzsch} $^a$ ~ and ~ {\bf Zhi-zhong Xing} $^b$
\thanks{E-mail: xingzz@mail.ihep.ac.cn}}
\address{$^a$ Physik-Department, Universit$\it\ddot{a}$t M$\it\ddot{u}$nchen,
Theresienstrasse 37A, 80333 Munich, Germany \\
$^b$ Institute of High Energy Physics and Theoretical Physics
Center for Science Facilities, \\ Chinese Academy of Sciences,
P.O. Box 918, Beijing 100049, China} \maketitle

\begin{abstract}
We propose two phenomenological scenarios of lepton mass matrices
and show that either of them can exactly give rise to
$\tan^2\theta^{}_{13} = m^{}_e/(m^{}_e + 2m^{}_\mu)$,
$\tan^2\theta^{}_{23} = m^{}_\mu/(m^{}_e + m^{}_\mu)$ and
$\tan^2\theta^{}_{12} = (m^{}_e m^{}_2 + 2m^{}_\mu m^{}_1)/(m^{}_e
m^{}_1 + 2m^{}_\mu m^{}_2)$ in the standard parametrization of
lepton flavor mixing. The third relation, together with current
experimental data, predicts a normal but weak hierarchy for the
neutrino mass spectrum. We also obtain $\theta^{}_{13} \approx
2.8^\circ$ for the smallest neutrino mixing angle and ${\cal J}
\approx 1.1\%$ for the Jarlskog invariant of leptonic CP violation,
which will soon be tested in the long-baseline reactor and
accelerator neutrino oscillation experiments. A seesaw realization
of both scenarios is briefly discussed.
\end{abstract}


\newpage

\section{Introduction}

In the standard parametrization of the Cabibbo-Kobayashi-Maskawa
(CKM) quark flavor mixing matrix $V$ \cite{PDG} a hierarchy exists
among its three mixing angles: $\vartheta^{}_{12} \sim \lambda$,
$\vartheta^{}_{23} \sim \lambda^2$, and $\vartheta^{}_{13} \sim
\lambda^4$ with $\lambda \approx 0.22$. This hierarchy might
intrinsically be related to the hierarchy of the quark masses:
$m^{}_u/m^{}_c \sim m^{}_c/m^{}_t \sim \lambda^4$ and $m^{}_d/m^{}_s
\sim m^{}_s/m^{}_b \sim \lambda^2$, since the CKM matrix $V$
actually measures a mismatch between the mass and flavor eigenstates
of both up- and down-type quarks. For the
Maki-Nakagawa-Sakata-Pontecorvo (MNSP) lepton flavor mixing matrix
$U$ \cite{MNS}, two of its three mixing angles are found to be
unexpectedly large: $\theta^{}_{12} \sim 34^\circ$ and
$\theta^{}_{23} \sim 45^\circ$. This phenomenon seems incompatible
with the mass hierarchy of the three charged leptons,
$m^{}_e/m^{}_\mu \sim \lambda^4/2$ and $m^{}_\mu/m^{}_\tau \sim
4\lambda^2/3$ \cite{PDG}. One might ascribe the largeness of
$\theta^{}_{12}$ and $\theta^{}_{23}$ to a very weak hierarchy of
the three neutrino masses (e.g., $m^{}_1 \sim m^{}_2 \sim m^{}_3$
\cite{FX96}). Another possibility of understanding the observed
pattern of neutrino mixing is to invoke a certain flavor symmetry
such that all three mixing angles are pure numbers in the symmetry
limit and their weak dependence on $m^{}_\alpha$ (for $\alpha = e,
\mu, \tau$) and $m^{}_i$ (for $i=1,2,3$) could result from the
details of the symmetry breaking \cite{Review}. One may also
speculate that one mixing angle is purely determined by an
underlying flavor symmetry, while the other two are more or less
associated with $m^{}_\alpha$ and $m^{}_i$, or vice versa.

In Refs. \cite{FX96,FX06}, we recommended a useful parametrization
of the MNSP matrix $U$:
\begin{eqnarray}
U & = & \left ( \matrix{ c^{}_l & s^{}_l   & 0 \cr -s^{}_l    &
c^{}_l   & 0 \cr 0   & 0 & 1 \cr } \right )  \left ( \matrix{
e^{-i\phi}  & 0 & 0 \cr 0   & c & s \cr 0   & -s    & c \cr } \right
)  \left ( \matrix{ c^{}_{\nu} & -s^{}_{\nu}  & 0 \cr s^{}_{\nu} &
c^{}_{\nu}   & 0 \cr
0   & 0 & 1 \cr } \right )  P \nonumber \\ \nonumber \\
& = & \left ( \matrix{ s^{}_l s^{}_{\nu} c + c^{}_l c^{}_{\nu}
e^{-i\phi} & s^{}_l c^{}_{\nu} c - c^{}_l s^{}_{\nu} e^{-i\phi} &
s^{}_l s \cr c^{}_l s^{}_{\nu} c - s^{}_l c^{}_{\nu} e^{-i\phi} &
c^{}_l c^{}_{\nu} c + s^{}_l s^{}_{\nu} e^{-i\phi} & c^{}_l s \cr -
s^{}_{\nu} s   & - c^{}_{\nu} s   & c \cr } \right ) P \;
\end{eqnarray}
with $c^{}_{l,\nu} \equiv \cos\theta^{}_{l,\nu}$, $s^{}_{l,\nu}
\equiv \sin\theta^{}_{l,\nu}$, $c \equiv \cos\theta$ and $s \equiv
\sin\theta$. $P = {\rm Diag} \left \{e^{i\rho}, e^{i\sigma}, 1
\right \}$ is the Majorana phase matrix. We conjectured that two
lepton mixing angles could be related to two lepton mass ratios in
the following way \cite{FX06}:
\begin{equation}
\tan^{}\theta^{}_l = \sqrt{\frac{m^{}_e}{m^{}_\mu}} ~ \; , ~~~~
\tan\theta^{}_\nu = \sqrt{\frac{m^{}_1}{m^{}_2}} ~ \; .
\end{equation}
This conjecture is reasonable because $\theta^{}_l$ and
$\theta^{}_\nu$ describe the $e$-$\mu$ mixing in the charged-lepton
sector and the $\nu^{}_e$-$\nu^{}_\mu$ mixing in the neutrino sector
respectively. Hence they can naturally vanish in the limit $m^{}_e
\rightarrow 0$ and $m^{}_1 \rightarrow 0$, consistent with the
experimental fact that $m^{}_e \ll m^{}_\mu$ and $m^{}_1 < m^{}_2$
hold. Note that the three mixing angles of $U$ in the standard
parametrization (i.e., $\theta^{}_{12}$, $\theta^{}_{13}$ and
$\theta^{}_{23}$) \cite{PDG} are approximately related to those in
Eq. (1):
\begin{equation}
\theta^{}_{12} \; \approx \; \theta^{}_\nu \; , ~~~~ \theta^{}_{23}
\; \approx \; \theta \; , ~~~~ \theta^{}_{13} \; \approx \;
\theta^{}_l \sin\theta \; ,
\end{equation}
since the solar and atmospheric neutrino oscillations can
approximately be decoupled from each other, Eq. (2) indicates that
the smallness of $\theta^{}_{13}$ is attributed to a very strong
charged-lepton mass hierarchy (i.e., $m^{}_e \ll m^{}_\mu$). The
large value of $\theta^{}_{12}$ is related to a relatively weak
neutrino mass hierarchy (i.e., $m^{}_1 < m^{}_2$).

In this paper we propose two simple scenarios of the charged-lepton
and neutrino mass matrices from which the phenomenological
conjecture made in Eq. (2) can exactly be derived. We show that both
scenarios can accommodate $\theta = \pi/4$ and $\phi = \pi/2$ in our
parametrization of $U$, leading to the predictions
$\tan^2\theta^{}_{13} = m^{}_e/(m^{}_e + 2m^{}_\mu)$,
$\tan^2\theta^{}_{23} = m^{}_\mu/(m^{}_e + m^{}_\mu)$ and
$\tan^2\theta^{}_{12} = (m^{}_e m^{}_2 + 2m^{}_\mu m^{}_1)/(m^{}_e
m^{}_1 + 2m^{}_\mu m^{}_2)$ in the standard parametrization of $U$.
We find that the third relation, together with current experimental
data, predicts a normal but weak hierarchy for the neutrino mass
spectrum: $m^{}_1 \approx 4.28 \times 10^{-3} ~ {\rm eV}$, $m^{}_2
\approx 9.74 \times 10^{-3} ~ {\rm eV}$ and $m^{}_3 \approx 4.92
\times 10^{-2} ~ {\rm eV}$. Furthermore, we obtain $\theta^{}_{13}
\approx 2.8^\circ$ for the smallest neutrino mixing angle and ${\cal
J} \approx 1.1\%$ for the Jarlskog invariant of leptonic CP
violation, which will soon be tested in the long-baseline reactor
and accelerator neutrino oscillation experiments. We discuss a
simple seesaw realization of these two scenarios.

\section{Two scenarios}

Our main goal is to reproduce the relations conjectured in Eq. (2)
and to predict $\theta = \pi/4$ for the MNSP matrix $U$. We find
that there are two simple possibilities of realizing this goal.
Their consequences on neutrino mixing and CP violation are exactly
the same.

\begin{center}
{\bf Scenario (A)}
\end{center}

Let us consider the following textures of charged-lepton and
neutrino mass matrices at the electroweak scale (i.e., $\Lambda =
M^{}_Z$):
\begin{equation}
M^{}_l = \left( \matrix{ 0 & A & 0 \cr ~A^*~ & B & 0 \cr 0 & 0 & ~C~
\cr} \right) \; , ~~~~ M^{}_\nu = \left( \matrix{ 0 & X & -X \cr X &
Y & Z \cr -X & Z & Y \cr} \right) \; .
\end{equation}
One may easily diagonalize the Hermitian matrix $M^{}_l$ by using
the unitary transformation $O^\dagger_l M^{}_l O^\prime_l = {\rm
Diag}\{m^{}_e, m^{}_\mu, m^{}_\tau\}$, where
\begin{eqnarray}
O^{}_l & = & \left( \matrix{ e^{i\phi^{}_l} & ~0~ & ~0 \cr 0 & 1 &
~0 \cr 0 & 0 & ~1 \cr} \right) \left( \matrix{ ~c^{}_l~ & -s^{}_l &
~0 \cr s^{}_l & ~~c^{}_l~~ & ~0 \cr 0 & 0 & ~1 \cr } \right) \; ,
\nonumber \\
O^\prime_l & = & \left( \matrix{ e^{i\phi^\prime_l} & ~0~ & ~0 \cr 0
& 1 & ~0 \cr 0 & 0 & ~1 \cr} \right) \left( \matrix{ c^{}_l & s^{}_l
& 0 \cr -s^{}_l & ~~c^{}_l~~ & 0 \cr 0 & 0 & 1 \cr } \right) \; ,
\end{eqnarray}
with $\phi^{}_l \equiv \arg(A) -\pi$, $\phi^\prime_l \equiv \arg(A)$
and $\tan^{}\theta^{}_l = \sqrt{m^{}_e/m^{}_\mu}~$. We have $|A| =
\sqrt{m^{}_e m^{}_\mu}~$, $B = m^{}_\mu - m^{}_e$ and $C =
m^{}_\tau$. Because $M^{}_\nu$ is symmetric, it can be diagonalized
by the transformation $O^\dagger_\nu M^{}_\nu O^*_\nu = {\rm
Diag}\{m^{}_1, m^{}_2, m^{}_3 \}$, where
\begin{eqnarray}
O^{}_\nu = \left( \matrix{ e^{i\phi^{}_\nu} & 0 & ~0 \cr 0 & c & ~s
\cr 0 & -s & ~c \cr} \right) \left( \matrix{ c^{}_\nu & -s^{}_\nu &
0 \cr s^{}_\nu & c^{}_\nu & 0 \cr 0 & 0 & 1 \cr } \right) P \; ,
\end{eqnarray}
with $\phi^{}_\nu \equiv \arg(X) - \pi$, $\theta = \pi/4$,
$\tan\theta^{}_\nu = \sqrt{m^{}_1/m^{}_2}~$ and $P = {\rm Diag}\{i,
1, 1\}$. We have $|X| = \sqrt{m^{}_1 m^{}_2/2}~$, $Y = (m^{}_3 +
m^{}_2 - m^{}_1)/2$ and $Z = (m^{}_3 - m^{}_2 + m^{}_1)/2$.

With the help of Eqs. (5) and (6), we are able to calculate the MNSP
matrix $U = O^\dagger_l O^{}_\nu$ with the definition of $\phi
\equiv \phi^{}_l - \phi^{}_\nu = \arg(A) - \arg(X)$. Then the
expression of $U$, which is compatible with our parametrization in
Eq. (1), can explicitly be given as
\begin{equation}
U = \left( \matrix{ \sqrt{\frac{m^{}_\mu}{m^{}_e + m^{}_\mu}} &
\sqrt{\frac{m^{}_e}{m^{}_e + m^{}_\mu}} & 0 \cr\cr
-\sqrt{\frac{m^{}_e}{m^{}_e + m^{}_\mu}} &
\sqrt{\frac{m^{}_\mu}{m^{}_e + m^{}_\mu}} & 0 \cr\cr 0 & 0 & 1 \cr}
\right) \left( \matrix{ e^{-i\phi} & 0 & 0 \cr\cr 0 &
\sqrt{\frac{1}{2}} & \sqrt{\frac{1}{2}} \cr\cr  0 &
-\sqrt{\frac{1}{2}} & \sqrt{\frac{1}{2}} \cr} \right) \left(
\matrix{ \sqrt{\frac{m^{}_2}{m^{}_1 + m^{}_2}} &
-\sqrt{\frac{m^{}_1}{m^{}_1 + m^{}_2}} & 0 \cr\cr
\sqrt{\frac{m^{}_1}{m^{}_1 + m^{}_2}} & \sqrt{\frac{m^{}_2}{m^{}_1 +
m^{}_2}} & 0 \cr\cr 0 & 0 & 1 \cr} \right) P \; .
\end{equation}
We assume $\phi = \pi/2$ for simplicity. This assumption is
consistent with the one made in the quark sector, where the
counterpart of $\phi$ is denoted as $\varphi$ \cite{FX97}. Indeed,
$\varphi = \pi/2$ is strongly favored by current experimental data
on the CKM unitarity triangle simply because $\varphi - \alpha
\approx 1.1^\circ$ holds \cite{Xing09,Antusch09} and $\alpha =
89.0^{+4.4^\circ}_{-4.2^\circ}$ has been measured \cite{Prell}).
Comparing Eq. (7) with the standard parametrization of the MNSP
matrix \cite{PDG},
\begin{eqnarray}
U & = & \left( \matrix{ 1 & 0 & 0 \cr 0 & c^{}_{23} & s^{}_{23} \cr
0 & -s^{}_{23} & c^{}_{23} \cr} \right) \left( \matrix{ c^{}_{13} &
0 & s^{}_{13} e^{-i\delta} \cr 0 & 1 & 0 \cr -s^{}_{13} e^{i\delta}
& 0 & c^{}_{13} \cr} \right) \left( \matrix{ c^{}_{12} & s^{}_{12} &
0 \cr -s^{}_{12} & c^{}_{12} & 0 \cr 0 & 0 & 1 \cr} \right) P
\nonumber \\
& = & \left( \matrix{ c^{}_{12} c^{}_{13} & s^{}_{12} c^{}_{13} &
s^{}_{13} e^{-i \delta} \cr -s^{}_{12} c^{}_{23} - c^{}_{12}
s^{}_{23} s^{}_{13} e^{i \delta} & c^{}_{12} c^{}_{23} - s^{}_{12}
s^{}_{23} s^{}_{13} e^{i \delta} & s^{}_{23} c^{}_{13} \cr s^{}_{12}
s^{}_{23} - c^{}_{12} c^{}_{23} s^{}_{13} e^{i \delta} & -c^{}_{12}
s^{}_{23} - s^{}_{12} c^{}_{23} s^{}_{13} e^{i \delta} & c^{}_{23}
c^{}_{13} \cr} \right) P \;
\end{eqnarray}
with $c^{}_{ij} \equiv \cos\theta^{}_{ij}$, $s^{}_{ij} \equiv
\sin\theta^{}_{ij}$ (for $ij = 12, 13, 23$) and $P = {\rm Diag}
\{e^{i\rho}, e^{i\sigma}, 1\}$, we immediately arrive at the
predictions
\begin{equation}
\tan^2\theta^{}_{12} = \frac{m^{}_e m^{}_2 + 2 m^{}_\mu
m^{}_1}{m^{}_e m^{}_1 + 2 m^{}_\mu m^{}_2} \; , ~~~~
\tan^2\theta^{}_{13} = \frac{m^{}_e}{m^{}_e + 2 m^{}_\mu} \; , ~~~~
\tan^2\theta^{}_{23} = \frac{m^{}_\mu}{m^{}_e + m^{}_\mu} \; ,
\end{equation}
together with $\rho = \pi/2$, $\sigma = 0$ and
\begin{equation}
\sin^2\delta = \frac{m^{}_1 m^{}_2 \left( m^{}_e + 2 m^{}_\mu
\right)^2}{\left( m^{}_e m^{}_1 + 2 m^{}_\mu m^{}_2 \right) \left(
m^{}_e m^{}_2 + 2 m^{}_\mu m^{}_1 \right)} \; .
\end{equation}
We have used $\phi = \pi/2$ in obtaining the results of
$\tan^2\theta^{}_{12}$ and $\sin^2\delta$. Given the strong mass
hierarchy $m^{}_e \ll m^{}_\mu$, Eqs. (9) and (10) can be written as
\begin{eqnarray}
\tan\theta^{}_{12} & \approx & \sqrt{\frac{m^{}_1}{m^{}_2}} \left( 1
+ \frac{m^{}_e}{m^{}_\mu} \frac{m^2_2 - m^2_1}{4 m^{}_1 m^{}_2}
\right) \; , ~~~~~ \tan\theta^{}_{23} \approx 1 - \frac{1}{2}
\frac{m^{}_e}{m^{}_\mu} \; ,
\nonumber \\
\tan\theta^{}_{13} & \approx & \sqrt{\frac{m^{}_e}{m^{}_\mu}} \left(
1 - \frac{1}{2} \frac{m^{}_e}{m^{}_\mu} \right) \; , ~~~~ \sin\delta
\approx 1 - \frac{m^{}_e}{m^{}_\mu} \frac{\left( m^{}_2 - m^{}_1
\right)^2}{4 m^{}_1 m^{}_2} \; .
\end{eqnarray}
Some comments on these results are in order.

(1) The neutrino mass ratio $m^{}_1/m^{}_2$ can be determined from
the experimental values of $m^{}_e/m^{}_\mu$ and
$\tan^2\theta^{}_{12}$. Eq. (9) yields
\begin{equation}
\frac{m^{}_1}{m^{}_2} = \frac{2 m^{}_\mu \tan^2\theta^{}_{12} -
m^{}_e}{2 m^{}_\mu - m^{}_e \tan^2\theta^{}_{12}} \approx
\tan^2\theta^{}_{12} \left( 1 - 2\frac{m^{}_e}{m^{}_\mu} \frac{\cos
2\theta^{}_{12}}{\sin^2 2\theta^{}_{12}} \right) \; .
\end{equation}
Because of $m^{}_e = 0.486570161 \pm 0.000000042$ MeV and $m^{}_\mu
= 102.7181359 \pm 0.0000092$ MeV at the scale $\Lambda = M^{}_Z$
\cite{XZZ08,Fusaoka}, we get $m^{}_e/m^{}_\mu \approx 0.004737$,
which is far below the error bar of $\tan^2\theta^{}_{12}$ extracted
from the present experimental data \cite{Data}. Hence it is good
enough to use the leading-order relation $m^{}_1/m^{}_2 \approx
\tan^2\theta^{}_{12}$ in our numerical calculation. Taking the
best-fit value $\sin^2\theta^{}_{12} = 0.304$ \cite{Data} for
example, we obtain $m^{}_1/m^{}_2 \approx 0.44$. This result,
together with the best-fit values $\Delta m^2_{21} \equiv m^2_2 -
m^2_1 = 7.65 \times 10^{-5} ~ {\rm eV}^2$ and $|\Delta m^2_{31}|
\equiv |m^2_3 - m^2_1| = 2.40 \times 10^{-3} ~ {\rm eV}^2$ obtained
from a global analysis of current neutrino oscillation data
\cite{Data}, allows us to determine the neutrino mass spectrum:
\begin{equation}
m^{}_1 \approx 4.28 \times 10^{-3} ~ {\rm eV} \; , ~~~~~ m^{}_2
\approx 9.74 \times 10^{-3} ~ {\rm eV} \; , ~~~~~ m^{}_3 \approx
4.92 \times 10^{-2} ~ {\rm eV} \; ,
\end{equation}
which exhibits a normal but weak hierarchy.

(2) The value of $\theta^{}_{13}$ and the deviation of
$\theta^{}_{23}$ from $\theta = \pi/4$ are both determined by the
ratio $m^{}_e/m^{}_\mu$, while the deviation of $\delta$ from $\phi
= \pi/2$ depends both on $m^{}_e/m^{}_\mu$ and $m^{}_1/m^{}_2$ in
this ansatz:
\begin{equation}
\theta^{}_{13} \approx 2.8^\circ \; , ~~~~~ \theta^{}_{23} \approx
44.9^\circ \; , ~~~~~ \delta \approx 87.6^\circ \; .
\end{equation}
In obtaining the value of $\delta$, we have used the values of
$m^{}_1$ and $m^{}_2$ in Eq. (13).

(3) The Jarlskog invariant of leptonic CP violation \cite{J} turns
out to be
\begin{equation}
{\cal J} = c^{}_l s^{}_l c^{}_\nu s^{}_\nu c s^2 \sin\phi =
c^{}_{12} s^{}_{12} c^2_{13} s^{}_{13} c^{}_{23} s^{}_{23}
\sin\delta = \frac{1}{2\sqrt{2}} \frac{\sqrt{m^{}_e m^{}_\mu m^{}_1
m^{}_2}}{(m^{}_e + m^{}_\mu) (m^{}_1 + m^{}_2)} \; .
\end{equation}
Taking the values of $m^{}_1$ and $m^{}_2$ given in Eq. (13), we
obtain ${\cal J} \approx 1.1\%$. This amount of CP violation can be
observed in the future long-baseline neutrino oscillation
experiments \cite{CP}, provided the terrestrial matter effects do
not mimic it \cite{Matter}.

With the help of Eq. (13) together with $m^{}_\tau =
1746.24^{+0.20}_{-0.19}$ MeV at $\Lambda = M^{}_Z$ \cite{XZZ08}, we
find the following textures of $M^{}_l$ and $M^{}_\nu$:
\begin{eqnarray}
M^{}_l & \approx & 1.00 m^{}_\tau \left( \matrix{ 0 & -0.0040
e^{+i\phi^{}_l} & 0 \cr -0.0040 e^{-i\phi^{}_l} & 0.058 & 0 \cr 0 &
0 & 1 \cr} \right) \; , \nonumber \\
M^{}_\nu & \approx & 0.56 m^{}_3 \left( \matrix{ 0 & -0.17
e^{i\phi^{}_\nu} & 0.17 e^{i\phi^{}_\nu} \cr -0.17 e^{i\phi^{}_\nu}
& 1 & 0.80 \cr 0.17 e^{i\phi^{}_\nu} & 0.80 & 1 \cr} \right) \; ,
\end{eqnarray}
where the phases $\phi^{}_l$ and $\phi^{}_\nu$ have been defined in
Eqs. (5) and (6), respectively. Eq. (16) clearly shows the weak
hierarchy of $M^{}_\nu$ in contrast with the strong hierarchy of
$M^{}_l$. It is therefore natural that two large flavor mixing
angles $\theta^{}_{12}$ and $\theta^{}_{23}$ stem from $M^{}_\nu$,
while the smallest flavor mixing angle $\theta^{}_{13}$ comes from
$M^{}_l$.

\begin{center}
{\bf Scenario (B)}
\end{center}

Now we consider the following ansatz of lepton mass matrices, whose
textures are essentially the interchanged ones of $M^{}_l$ and
$M^{}_\nu$ given in Eq. (4), at the electroweak scale:
\begin{equation}
\tilde{M}^{}_l = \left( \matrix{ 0 & \tilde{A} & -\tilde{A} \cr
\tilde{A}^* & \tilde{B} & \tilde{C} \cr -\tilde{A}^* & \tilde{C} &
\tilde{B} \cr} \right) \; , ~~~~ \tilde{M}^{}_\nu = \left( \matrix{
0 & \tilde{X} & 0 \cr ~\tilde{X}~ & \tilde{Y} & 0 \cr 0 & 0 &
~\tilde{Z}~ \cr} \right) \; .
\end{equation}
The Hermitian matrix $\tilde{M}^{}_l$ is diagonalized by using the
unitary transformation $\tilde{O}^\dagger_l \tilde{M}^{}_l
\tilde{O}^\prime_l = {\rm Diag}\{m^{}_e, m^{}_\mu, m^{}_\tau\}$,
where
\begin{eqnarray}
\tilde{O}^{}_l & = & \left( \matrix{ e^{i\phi^{}_l} & 0 & ~0 \cr 0 &
c & -s \cr 0 & s & c \cr} \right) \left( \matrix{ c^{}_l & -s^{}_l &
~0 \cr s^{}_l & c^{}_l & ~0 \cr 0 & 0 & ~1 \cr } \right)
\; , \nonumber \\
\tilde{O}^\prime_l & = & \left( \matrix{ e^{i\phi^\prime_l} & 0 & ~0
\cr 0 & c & -s \cr 0 & s & c \cr} \right) \left( \matrix{ c^{}_l &
s^{}_l & ~0 \cr -s^{}_l & c^{}_l & ~0 \cr 0 & 0 & ~1 \cr } \right)
\; ,
\end{eqnarray}
with $\phi^{}_l \equiv \arg(\tilde{A}) -\pi$, $\phi^\prime_l \equiv
\arg(\tilde{A})$, $\theta = -\pi/4$ and $\tan^{}\theta^{}_l =
\sqrt{m^{}_e/m^{}_\mu}~$. In addition, we have $|\tilde{A}| =
\sqrt{m^{}_e m^{}_\mu/2}~$, $\tilde{B} = (m^{}_\tau + m^{}_\mu -
m^{}_e)/2$ and $\tilde{C} = (m^{}_\tau - m^{}_\mu + m^{}_e)/2$.
Since $\tilde{M}^{}_\nu$ is symmetric, it can be diagonalized by
means of the transformation $\tilde{O}^\dagger_\nu \tilde{M}^{}_\nu
\tilde{O}^*_\nu = {\rm Diag}\{m^{}_1, m^{}_2, m^{}_3 \}$, where
\begin{eqnarray}
\tilde{O}^{}_\nu = \left( \matrix{ e^{i\phi^{}_\nu} & 0 & ~0 \cr 0 &
1 & ~0 \cr 0 & 0 & ~1 \cr} \right) \left( \matrix{ c^{}_\nu &
-s^{}_\nu & 0 \cr s^{}_\nu & c^{}_\nu & 0 \cr 0 & 0 & 1 \cr }
\right) P \; ,
\end{eqnarray}
with $\phi^{}_\nu \equiv \arg(\tilde{X}) - \pi$, $\tan\theta^{}_\nu
= \sqrt{m^{}_1/m^{}_2}~$ and $P = {\rm Diag}\{i, 1, 1\}$.
Furthermore, we have $|\tilde{X}| = \sqrt{m^{}_1 m^{}_2}~$,
$\tilde{Y} = m^{}_2 - m^{}_1$ and $\tilde{Z} = m^{}_3$.

We are then able to calculate the MNSP matrix $\tilde{U} =
\tilde{O}^\dagger_l \tilde{O}^{}_\nu$ with the definition of $\phi
\equiv \phi^{}_l - \phi^{}_\nu = \arg(\tilde{A}) - \arg(\tilde{X})$.
The result of $\tilde{U}$, after a redefinition of the phases of the
$\tau$ and $\nu^{}_3$ fields (i.e., $\tau \rightarrow -\tau$ and
$\nu^{}_3 \rightarrow -\nu^{}_3$ in order to absorb the minus sign
coming from $\theta = -\pi/4$), is $\tilde{U} = U \tilde{P}$, where
$U$ has been given in Eq. (7) and $\tilde{P} = {\rm Diag}\{i, 1,
-1\}$ holds. Given $\phi = \pi/2$, the predictions of the present
scenario for $\theta^{}_{12}$, $\theta^{}_{13}$, $\theta^{}_{23}$
and $\delta$ in the standard parametrization are exactly the same as
those obtained in Eqs. (9) and (10). Hence the scenarios of lepton
mass matrices proposed in Eqs. (4) and (17) are phenomenologically
identical and indistinguishable.

To illustrate the textures of $\tilde{M}^{}_l$ and
$\tilde{M}^{}_\nu$, we adopt the values of three neutrino masses in
Eq. (13) and the central values of three charged-lepton masses at
$\Lambda = M^{}_Z$. We obtain
\begin{eqnarray}
\tilde{M}^{}_l & \approx & 0.53 m^{}_\tau \left( \matrix{ 0 &
-0.0054 e^{+i\phi^{}_l} & 0.0054 e^{+i\phi^{}_l} \cr -0.0054
e^{-i\phi^{}_l} & 1 & 0.89 \cr 0.0054 e^{-i\phi^{}_l} &
0.89 & 1 \cr} \right) \; , \nonumber \\
\tilde{M}^{}_\nu & \approx & 1.00 m^{}_3 \left( \matrix{ 0 & -0.13
e^{i\phi^{}_\nu} & 0 \cr -0.13 e^{i\phi^{}_\nu} & 0.11 & 0 \cr 0 & 0
& 1 \cr} \right) \; ,
\end{eqnarray}
where the phases $\phi^{}_l$ and $\phi^{}_\nu$ have been defined in
Eqs. (5) and (6), respectively. A comparison between Eqs. (16) and
(20) shows the slight changes of the structural hierarchies in both
the charged-lepton and neutrino sectors.

\section{Seesaw mechanism}

The simple neutrino mass texture discussed above implies that it can
be derived from the canonical seesaw mechanism \cite{SS}, which
attributes the small masses of the three known neutrinos to the
existence of some heavy degrees of freedom far above the electroweak
scale $\Lambda = M^{}_Z$. Introducing three right-handed neutrinos
into the standard model, one can write down the gauge-invariant
Lagrangian associated with the lepton masses. After spontaneous
gauge symmetry breaking the resultant lepton mass terms are
\begin{eqnarray}
-{\cal L}^{}_{\rm lepton} = \overline{l^{}_{\rm L}} M^{}_l l^{}_{\rm
R} + \overline{\nu^{}_{\rm L}} M^{}_{\rm D} N^{}_{\rm R} +
\frac{1}{2} \overline{N^c_{\rm R}} M^{}_{\rm R} N^{}_{\rm R} + {\rm
h.c.} \; ,
\end{eqnarray}
where $M^{}_l$ and $M^{}_{\rm D}$ stand respectively for the mass
matrices of charged leptons and Dirac neutrinos, while $M^{}_{\rm
R}$ is a symmetric Majorana mass matrix of right-handed neutrinos.
As the Majorana mass term is not subject to the electroweak symmetry
breaking, the scale of $M^{}_{\rm R}$ can be much higher than those
of $M^{}_l$ and $M^{}_{\rm D}$ characterized by $M^{}_Z$. Hence the
effective mass matrix of the three active neutrinos is given by the
well-known seesaw relation \cite{SS}
\begin{equation}
M^{}_\nu \approx M^{}_{\rm D} M^{-1}_{\rm R} M^T_{\rm D} \;
\end{equation}
to a good degree of accuracy. Although the small mass scale of
$M^{}_\nu$ is qualitatively ascribed to the smallness of $M^{}_{\rm
D}/M^{}_{\rm R}$ in Eq. (22), this seesaw picture itself has no
quantitative predictability because the flavor structures of
$M^{}_{\rm D}$ and $M^{}_{\rm R}$ are entirely unspecified. To
generate large flavor mixing angles in the neutrino sector, extra
assumptions must be made so as to specify the textures of $M^{}_{\rm
D}$ and $M^{}_{\rm R}$ in the seesaw framework. Instead of going
into any details of model building, here we give some brief
discussions about the seesaw-invariant property of neutrino mass
matrices for our scenarios.

We first discuss a seesaw realization of the texture of $M^{}_\nu$
proposed in Eq. (4). We assume that $M^{}_{\rm D}$ and $M^{}_{\rm
R}$ have a similar texture which might result from a certain
(underlying) flavor symmetry:
\begin{equation}
M^{}_{\rm D} = \left( \matrix{ 0 & x & -x \cr x & y & z \cr -x & z &
y \cr} \right) \; , ~~~~ M^{}_{\rm R} = \left( \matrix{ 0 & {\bf X}
& -{\bf X} \cr {\bf X} & {\bf Y} & {\bf Z} \cr -{\bf X} & {\bf Z} &
{\bf Y} \cr} \right) \; ,
\end{equation}
With the help of Eq. (22), one can easily show that $M^{}_\nu$ takes
the same texture as $M^{}_{\rm D}$ and $M^{}_{\rm R}$
\cite{Matsuda}. Thus it is consistent with scenario (A). The matrix
elements of $M^{}_\nu$ turn out to be
\begin{eqnarray}
X & = & \displaystyle\frac{x^2}{\bf X} \; , \nonumber \\
Y & = & \displaystyle\frac{\left( y + z\right)^2}{2\left( {\bf Y} +
{\bf Z}\right)} + \frac{x \left( y - z\right)}{\bf X} - \frac{x^2
\left( {\bf Y} - {\bf Z} \right)}{2 {\bf X}^2}  \; , \nonumber \\
Z & = & \displaystyle\frac{\left( y + z\right)^2}{2\left( {\bf Y} +
{\bf Z}\right)} - \frac{x \left( y - z\right)}{\bf X} + \frac{x^2
\left( {\bf Y} - {\bf Z} \right)}{2 {\bf X}^2}  \; .
\end{eqnarray}
Such a seesaw-invariant texture is interesting since its matrix
elements can be expressed in terms of the corresponding mass
eigenvalues.

Now we examine whether the following texture of $M^{}_{\rm D}$ and
$M^{}_{\rm R}$ is seesaw-invariant for scenario (B):
\begin{equation}
\tilde{M}^{}_{\rm D} = \left( \matrix{ 0 & \tilde{x} & 0 \cr
\tilde{x} & \tilde{y} & 0 \cr 0 & 0 & \tilde{z} \cr} \right) \; ,
~~~~ \tilde{M}^{}_{\rm R} = \left( \matrix{ 0 & \tilde{\bf X} & 0
\cr \tilde{\bf X} & \tilde{\bf Y} & 0 \cr 0 & 0 & \tilde{\bf Z} \cr}
\right) \; .
\end{equation}
By using the seesaw formula $\tilde{M}^{}_\nu \approx
\tilde{M}^{}_{\rm D} \tilde{M}^{-1}_{\rm R} \tilde{M}^T_{\rm D}$, we
find that $\tilde{M}^{}_\nu$ has the same texture which is
consistent with scenario (B) proposed in Eq. (17) \cite{Nishiura}.
The non-vanishing matrix elements of $\tilde{M}^{}_\nu$ are given as
\begin{equation}
\tilde{X} = \frac{\tilde{x}^2}{\tilde{\bf X}} \; , ~~~~~~ \tilde{Y}
= \frac{2\tilde{x}\tilde{y}}{\tilde{\bf X}} - \frac{\tilde{x}^2
\tilde{\bf Y}}{\tilde{\bf X}^2} \; , ~~~~~~ \tilde{Z} =
\frac{\tilde{z}^2}{\tilde{\bf Z}} \; .
\end{equation}
One may express the matrix elements of $\tilde{M}^{}_{\rm D}$,
$\tilde{M}^{}_{\rm R}$ or $\tilde{M}^{}_\nu$ in terms of the
corresponding mass eigenvalues, but there remain some unknown
degrees of freedom (e.g., the phases in each mass matrix).

We conclude with two brief comments on scenarios (A) and (B).

(1) The texture of $M^{}_{\rm D}$, $M^{}_{\rm R}$ and $M^{}_\nu$ in
scenario (A) has the following $Z^{}_2$-like symmetry:
\begin{equation}
T M^{}_{\rm D} T = M^{}_{\rm D} \; , ~~~~ T M^{}_{\rm R} T =
M^{}_{\rm R} \; , ~~~~ T M^{}_\nu T = M^{}_\nu \; ,
\end{equation}
where
\begin{equation}
T = \left( \matrix{ -1 & 0 & ~0 \cr 0 & 0 & ~1 \cr 0 & 1 & ~0 \cr}
\right) \; .
\end{equation}
It is therefore possible to obtain such a texture by introducing
several Higgs doublets into the Lagrangian and imposing a global
symmetry like the one in Eq. (27) on the neutrino mass terms
\cite{Grimus,Koide}. An implementation of the global $U(1) \times
Z^{}_2$ symmetry may allow one to obtain the texture of $M^{}_{\rm
D}$, $M^{}_{\rm R}$ and $M^{}_\nu$ in scenario (B) \cite{Branco}.

(2) The seesaw scale $\Lambda^{}_{\rm SS}$ is usually much higher
than the electroweak scale $M^{}_Z$, but the texture of $M^{}_\nu$
(or $\tilde{M}^{}_\nu$) is expected to be essentially stable against
radiative corrections from $\Lambda^{}_{\rm SS}$ down to $M^{}_Z$
\cite{RGE}. Because the neutrino mass spectrum has a normal
hierarchy in either scenario (A) or scenario (B), radiative
corrections to flavor mixing angles and CP-violating phases are
actually negligible in most cases \cite{RGE}.

\section{Summary}

We have proposed two simple scenarios of the lepton mass matrices
from which the interesting relations $\tan\theta^{}_l =
\sqrt{m^{}_e/m^{}_\mu}~$ and $\tan\theta^{}_\nu =
\sqrt{m^{}_1/m^{}_2}~$ can be derived
\footnote{Note that the approximate relations of this type can also
be derived from the four-zero textures of lepton (or quark) mass
matrices \cite{FX99}.}.
Both scenarios can be realized in a seesaw framework. They predict
$\tan^2\theta^{}_{13} = m^{}_e/(m^{}_e + 2m^{}_\mu)$,
$\tan^2\theta^{}_{23} = m^{}_\mu/(m^{}_e + m^{}_\mu)$ and
$\tan^2\theta^{}_{12} = (m^{}_e m^{}_2 + 2m^{}_\mu m^{}_1)/(m^{}_e
m^{}_1 + 2m^{}_\mu m^{}_2)$ in the standard parametrization of the
MNSP matrix. We find that the third relation, together with current
experimental data, allows us to obtain a normal but weak hierarchy
for the neutrino mass spectrum: $m^{}_1 \approx 4.28 \times 10^{-3}
~ {\rm eV}$, $m^{}_2 \approx 9.74 \times 10^{-3} ~ {\rm eV}$ and
$m^{}_3 \approx 4.92 \times 10^{-2} ~ {\rm eV}$. Our predictions
$\theta^{}_{13} \approx 2.8^\circ$ for the smallest neutrino mixing
angle and ${\cal J} \approx 1.1\%$ for the Jarlskog invariant of CP
violation will soon be tested in the long-baseline accelerator and
reactor neutrino oscillation experiments \cite{CP}, such as the Daya
Bay experiment \cite{DB}.

\vspace{1cm}

{\it Acknowledgement:} The work of Z.Z.X. is supported in part by
the National Natural Science Foundation of China under grant No.
10425522 and No. 10875131.

\end{document}